\documentclass[11pt,a4paper]{article}
\pdfoutput=1
\usepackage{jcappub}

\newcommand{\PRE}[1]{{#1}} 

\newcommand{\be}{\begin{equation}}
\newcommand{\ee}{\end{equation}}
\newcommand{\bea}{\begin{eqnarray}}
\newcommand{\eea}{\end{eqnarray}}

\def\gev{\text{ GeV}}
\def\mev{\text{ MeV}}

\def\pb{\text{ pb}}

\def\kT{\text{ kT}}
\def\sr{\text{ sr}}
\def\cm{\text{ cm}}
\def\m{\text{ m}}
\def\km{\text{ km}}

\def\s{\text{ s}}
\def\yr{\text{ yr}}

\def\beq{\begin{eqnarray}}
\def\eeq{\end{eqnarray}}
\def\bea{\begin{eqnarray}}
\def\eea{\end{eqnarray}}

\def\sigmaSD{\sigma_{\rm SD}}

\newcommand{\gsim}{\lower.7ex\hbox{$\;\stackrel{\textstyle>}{\sim}\;$}}
\newcommand{\lsim}{\lower.7ex\hbox{$\;\stackrel{\textstyle<}{\sim}\;$}}

\usepackage{enumerate}
\include{epsf}

\title{
\textsc{Dark Matter Searches for Monoenergetic Neutrinos Arising from
Stopped Meson Decay in the Sun}
\PRE{\vspace*{0.1in}}
}

\author[a]{Carsten Rott}
\author[a]{Seongjin In}
\author[b]{Jason Kumar}
\author[c,d]{David Yaylali}

\affiliation[a]{\mbox{Department of Physics, Sungkyunkwan University, Suwon 440-746, Korea}}
\affiliation[b]{\mbox{Department of Physics \& Astronomy, University of
Hawai'i, Honolulu, HI 96822, USA}}
\affiliation[c]{\mbox{Department of Physics, University of Arizona, Tucson, AZ 85721, USA}}
\affiliation[d]{\mbox{Department of Physics, University of Maryland, College Park, MD 20742, USA}
\PRE{\vspace*{.1in}}
}

\emailAdd{rott@skku.edu}
\emailAdd{seongjin.in@gmail.com}
\emailAdd{jkumar@hawaii.edu}
\emailAdd{yaylali@email.arizona.edu}

\abstract{
Dark matter can be gravitationally captured by the Sun after scattering off solar nuclei. Annihilations of the dark matter trapped and accumulated in the centre of the Sun could
result in one of the most detectable and recognizable signals for dark matter. Searches for high-energy neutrinos produced in the decay of annihilation products have yielded extremely competitive
constraints on the spin-dependent scattering cross sections of dark matter with nuclei. Recently, the \textit{low energy} neutrino signal arising from dark-matter annihilation to quarks which then hadronize and shower has been suggested as a competitive and complementary search strategy.  These high-multiplicity hadronic showers give rise to a large amount of pions which will come to rest in the Sun and decay, leading to a unique sub-GeV neutrino signal.  We here improve on previous works by considering the monoenergetic neutrino
signal arising from both pion and kaon
decay. We consider searches at liquid scintillation, liquid argon, and water Cherenkov detectors and find very competitive sensitivities for few-GeV dark matter masses.
}


\begin{document}

\begin{flushright}
{\large \tt 
UH511-1249-15  CETUP2015-020}  
\end{flushright}

\maketitle
\flushbottom
\date{\today}


\section{Introduction}

Dark matter which scatters off solar nuclei can be gravitationally
captured by the Sun.  As a result, the dark matter density within the Sun can be much higher than the local density of the dark matter halo, potentially providing us an opportune region in which to search for experimental signatures of dark matter.  Indeed, one of the main indirect-detection search strategies is the search for neutrinos originating from the Sun which are produced through the annihilation of this dark matter~\cite{Silk:1985ax,Press:1985ug,Krauss:1985ks}.

A key aspect of this strategy is the search for energetic neutrinos arising from the prompt decays of the some dark matter (DM) annihilation products. For GeV-scale dark matter, these neutrinos would have energies well above the solar and atmospheric neutrino backgrounds at Earth-based detectors, providing a clear signal of DM annihilation. Experimental searches for these energetic neutrinos have resulted in tight bounds
on DM scattering on nucleons~\cite{Aartsen:2012kia,Choi:2015ara,Adrian-Martinez:2013ayv},
the processes that initiates the capture in the Sun.
It was long thought, however, that a class of models which could {\it not} be probed by this search
strategy was models in which the dark matter primarily annihilates to $u$,
$d$, and $s$ quarks.  The reason is that these quarks will quickly hadronize and the hadrons will interact and lose energy in the Sun before decaying,
thus yielding few high-energy neutrinos.   Not only would the low-energy neutrinos produced through these hadronic decays be competing against large backgrounds, but detection of these neutrinos becomes difficult as the cross section for neutrino--nucleus scattering in the detector decreases with decreasing energy.  As a result, this annihilation channel was largely ignored.

In~\cite{Rott:2012qb,Bernal:2012qh}, however, it was pointed out that the stopping of light hadrons in the dense solar medium produces a cascade of
other light hadrons, resulting in an abundance of stopped $\pi^+$ whose decays yield a well determined spectrum of sub-GeV neutrinos.  In essence, when looking at the light-quark annihilation channels one trades a hard neutrino spectrum for a spectrum which is softer, but with a larger amplitude and very distinctive spectral features.

In this work, we will refine the analysis put forth in~\cite{Rott:2012qb} by 
performing a more complete analysis of the hadronic processes in the Sun after DM annihilation.
We will show that competitive sensitivities can be obtained by focusing on the \textit{monoenergetic} $\nu_\mu$ at 29.8~MeV and 235.6~MeV, which are produced by the decay at rest of $\pi^+$ and $K^+$, respectively.  The monoenergetic neutrinos can oscillate into electron neutrinos, which produce line-signals in neutrino detectors with good energy resolution, such as liquid scintillation (LS) detectors and liquid argon time projection chambers (LArTPCs)~\cite{Kumar:2015nja}.  The line signal yields a large signal-to-background ratio, resulting in excellent detection prospects for low-mass dark matter. As a result, these detectors can probe classes of models which were previously thought beyond their reach. We discuss the sensitivity which can be obtained by generic LS or LArTPC detectors, but with a specific focus on two benchmark examples, KamLAND and DUNE. We compare these results with the sensitivity of large water Cherenkov (WC) detectors such as Super-Kamiokande~\cite{Fukuda:2002uc} and the proposed upgrade Hyper-Kamiokande~\cite{Abe:2015zbg,Abe:2011ts}.

It is worthwhile briefly discussing the theoretical prejudice against models in which dark matter annihilates to
light quarks.  If dark matter (denoted as $X$) is a Majorana fermion and if flavor violation is minimal, then the branching fraction
for dark-matter annihilation to light quarks ($XX \rightarrow \bar q q$) is suppressed.  Because the initial state consists of identical fermions,
the wavefunction must be totally antisymmetric, implying that $s$-wave annihilation can only take place from a $J=0$
state.  Angular momentum conservation then implies that the outgoing $\bar q$ and $q$ must have the same helicity, and
thus arise from different Weyl spinors.  Minimal flavor violation would require that any such Weyl spinor mixing be proportional to
the quark mass, suppressing annihilation to $\bar q q$  for $q=u,d,s$.  But this argument relies crucially on the assumptions of Majorana
fermion dark matter and minimal flavor violation; if either assumption fails, then the branching fraction for dark matter
annihilation to light quarks can easily be ${\cal O}(1)$\cite{Kumar:2015nja,Kumar:2013iva}.  Models of this type can be probed by
searches for low-energy neutrinos arising from dark-matter annihilation in the Sun, as described in this work.

This paper is structured as follows.  In section II, we describe the details of the analysis, including the
expected signal and background event rates.  In section III we present the expected sensitivities for KamLAND, DUNE,
Super-Kamiokande (Super-K) and Hyper-Kamiokande (Hyper-K). We conclude in section IV.

\section{Analysis}

We first provide an overview of the framework of the analysis.
A dark matter particle will be gravitationally captured by the Sun if, in scattering against solar
nuclei, it falls below the escape velocity. We refer the interested reader to
Refs.~\cite{Gould:1987ir,Gould:1991hx} for the standard capture calculation.
The capture of dark matter in the Sun is rather insensitive to the
underlying dark-matter velocity distribution and astrophysical conditions~\cite{Choi:2013eda,Danninger:2014xza}.
For standard scenarios the Sun is expected to be in equilibrium, meaning that the annihilation
rate ($\Gamma_{\text{A}}$) is related to the capture rate ($\Gamma_{\text{C}}$) by $\Gamma_{\text{A}} = (1/2)\Gamma_{\text{C}}$.  Thus a constraint on the flux of neutrinos from the DM annihilation products provides a constraint on the DM--nucleus scattering cross section.

Annihilation processes to light quarks ($XX \rightarrow \bar q q$) will then produce energetic light hadrons.  While very short-lived hadrons will decay promptly, the longer-lived \textit{pseudo-stable} hadrons will interact with the dense nuclear plasma in the Sun's core before decaying, losing energy in the process.  For dark-matter masses in the GeV range, many of these longer-lived hadrons will be energetic enough to create a multitude of secondary hadrons from these interactions, creating a large shower of additional light pseudo-stable particles.  As energy is dispersed into the shower, low-energy elastic scattering processes with the solar plasma take over and the longer-lived particles  --- such as pions and kaons --- will eventually come to rest.
Some of these hadrons will then decay, at rest, to a number of neutrinos with well determined energy spectra.

Our interest will be in $\pi^+$ and $K^+$, which are produced at a rate which is directly proportional to the annihilation rate; decay then proceeds via the processes $\pi^+, K^+ \rightarrow \nu_\mu \mu^+$, yielding a flux of monoenergetic neutrinos.  Other particles produced in the hadronic shower will not contribute significantly to any signal, either because their multiplicity is very low, their decays do not yield monoenergetic neutrinos of interest, or they interact before decaying. For example $\pi^0$ will decay largely to two
photons in a very fast electromagnetic process, while $\pi^-$ will be Coulomb-captured by nuclei in the Sun via processes which do not produce neutrinos \cite{Ponomarev:1973ya}.

The process $\pi^+ \rightarrow \nu_\mu \mu^+$ occurs with a branching fraction of nearly 100\%, while the
process $K^+ \rightarrow \nu_\mu \mu^+$ occurs with branching fraction $\sim 64\%$.
The energy of the produced monoenergetic neutrino is given by
\bea
E &=& {m_{\pi^\pm, K^\pm}^2 - m_\mu^2 \over 2m_{\pi^\pm, K^\pm} } .
\eea
For $\pi^+$ ($K^+$) decay, the energy of the monoenergetic neutrino will be $E \sim 29.8\mev$ ($235.6\mev$).
The monoenergetic $\nu_\mu$ will oscillate to all three neutrino flavors, but we will be interested in $\nu_e$ which
interact through a charged-current interaction at the detector.
For such an interaction, almost all of the energy will
be deposited in the detector~\cite{Kumar:2009ws,Kumar:2011hi}, allowing for a complete reconstruction of the line signal.
For the final-state electron to emerge with an energy near that of the incoming neutrino, the difference in binding energy
between the initial-state nucleus and final-state nucleus must be small compared to the neutrino energy.  This requirement
makes argon an ideal target material, but oxygen and carbon are also good targets for the $235.6\mev$ line.

Note, the $\mu^+$ produced by meson decay will also stop in the Sun before itself decaying ($\mu^+ \rightarrow \bar \nu_\mu \nu_e e^+$),
yielding a continuum
$\bar \nu_\mu$ and $\nu_e$ Michel spectrum~\cite{Michel:1949qe,Kinoshita:1958ru}.  Electron anti-neutrinos produced
in oscillations from this continuum spectrum can be efficiently detected via inverse beta decay reactions and have been
subject to previous studies~\cite{Rott:2012qb,Bernal:2012qh}.  We find, however, that this signal is
less distinctive and not as competitive as the line feature studied here and hence we will not consider it further.

\subsection{The Neutrino Flux From Stopped Meson Decay in the Sun}

The capture rate may be expressed as
$\Gamma_{\text{C}} = C_0^{\text{SD}}(m_X) \times \sigmaSD^p \times [(\rho_X / \rho_\odot) (\bar v / 270 \km /\text{s})^{-1}]$~\cite{Gould:1987ir},
where values of $C_0^{\text{SD}} (m_X)$ are given, for example, in~\cite{Gao:2011bq,Kumar:2012uh}.  Here, $\sigmaSD^p$ is the
the dark-matter--proton spin-dependent scattering cross section, $\rho_X$ is the dark-matter density,
$\rho_\odot = 0.3\gev/\text{cm}^3$, and $\bar v$ is the dark-matter velocity dispersion of a Maxwell-Boltzmann distribution.
Note, we assume that dark-matter--nucleon scattering is spin-dependent, because spin-independent scattering is
already tightly constrained by direct-detection experiments.

We will consider the range of dark matter mass for which the effects of evaporation are negligible
($m_X \gtrsim 4\gev$)~\cite{WIMPevaporation}.  For dark matter with $m_X \sim 10\gev$ and an
annihilation cross section $\langle \sigma v \rangle \sim 1 \pb$, the Sun will be in equilibrium if
$\sigmaSD^p \gtrsim 3 \times 10^{-7} \pb$~\cite{Kumar:2012uh}.  We will thus assume that the Sun is in
equilibrium for the remainder of this work, implying $\Gamma_{\text{A}} = (1/2)\Gamma_{\text{C}}$.

Each dark-matter annihilation will result in the energy of the dark matter particles being partitioned
among a variety of Standard Model particles.
We will define $r_{\pi,K}$ as the fraction of the center-of-mass energy of each annihilation which
goes into either stopped $\pi^+$ or $K^+$; if $n_{\pi, K}$ is the average number of stopped $\pi^+, K^+$ produced
per annihilation, then
\bea
r_{\pi, K} &\equiv& {m_{\pi, K} n_{\pi, K} \over 2m_X} .
\eea
The flux of monoenergetic electron neutrinos at Earth arising from stopped mesons in the Sun is then given by
\bea
{d^2 \Phi_{\pi, K} \over dE d\Omega} &=&
{(F_{\nu_e})(1/2) \Gamma_{\text{C}} \over 4\pi r_\oplus^2} \left({2 m_X r_{\pi, K} \over m_{\pi, K} } \right)
\delta(E-E_0) \delta(\Omega) ,
\eea
where $r_\oplus \sim 1.5 \times 10^{11} \m$ is the Earth-Sun distance and $F_{\nu_e}$ is the fraction of the injected
monoenergetic $\nu_\mu$ which have oscillated to $\nu_e$ by the time they reach the detector.  $E_0$ is the
energy of the monoenergetic neutrino, and the $\delta (\Omega)$ factor enforces the condition that the neutrino
flux emanates from the Sun.

The $r_{\pi, K}$ are obtained by simulating the showering and hadronization effects for the annihilation processes in \textsc{Pythia~8.2}~\cite{Sjostrand:2014zea}, and then simulating the propagation of the annihilation products in
the Sun using \textsc{GEANT4}~\cite{Agostinelli:2002hh}.  By simulating the initial dark-matter annihilation process in \textsc{Pythia}, we find the distributions of multiplicity and energy of various hadrons produced.  Given the energy and lifetime of each hadron species, we can then determine which hadrons have a path-length which is large compared to the inter-nuclear spacing in the core of the Sun before decay.  The hadrons with shorter path-lengths are then decayed within \textsc{Pythia}, before injection into the \textsc{GEANT4} intrasolar-propagation simulation.

\begin{figure}[t]
   \includegraphics[scale=0.40]{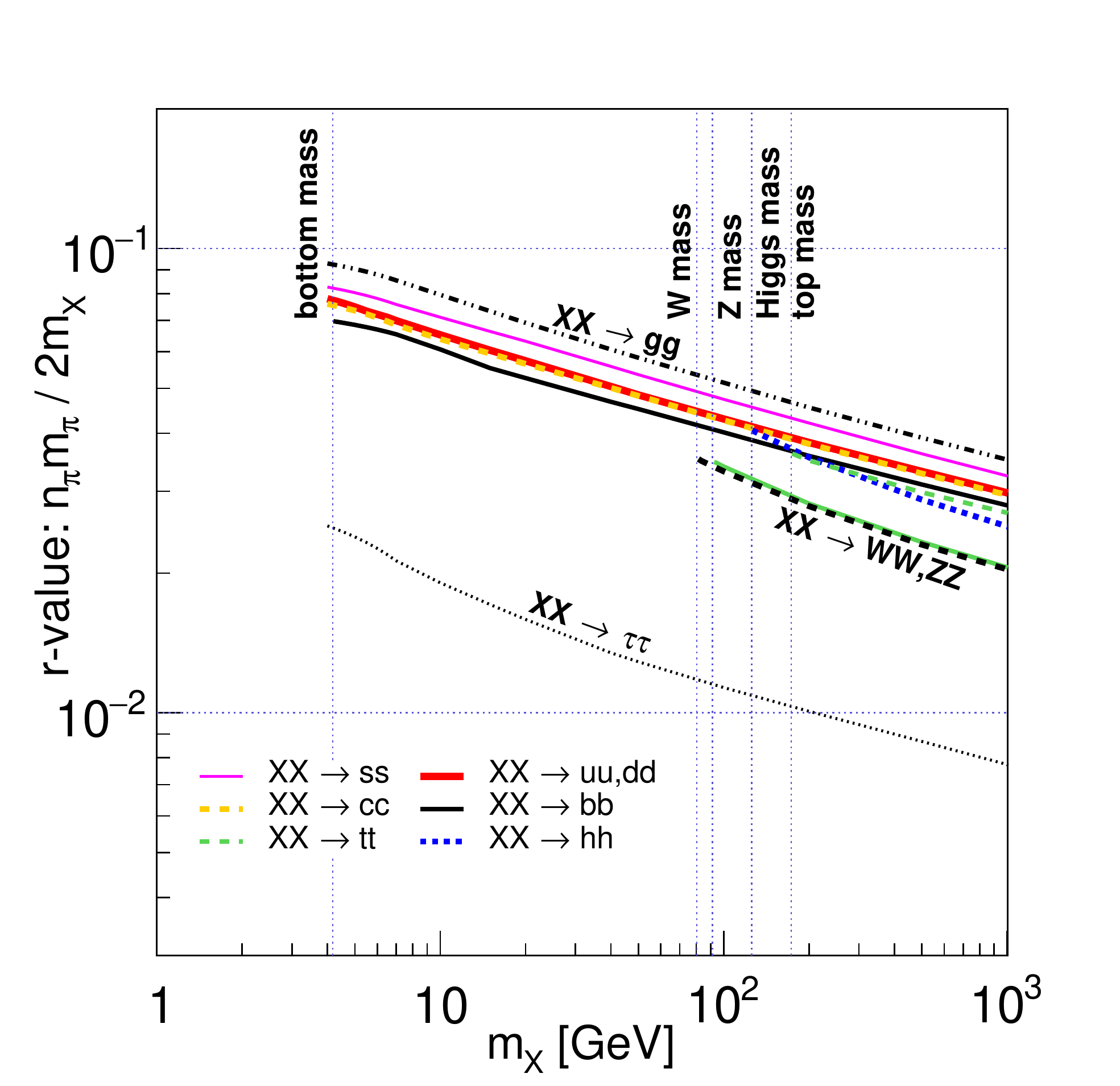}
   \includegraphics[scale=0.40]{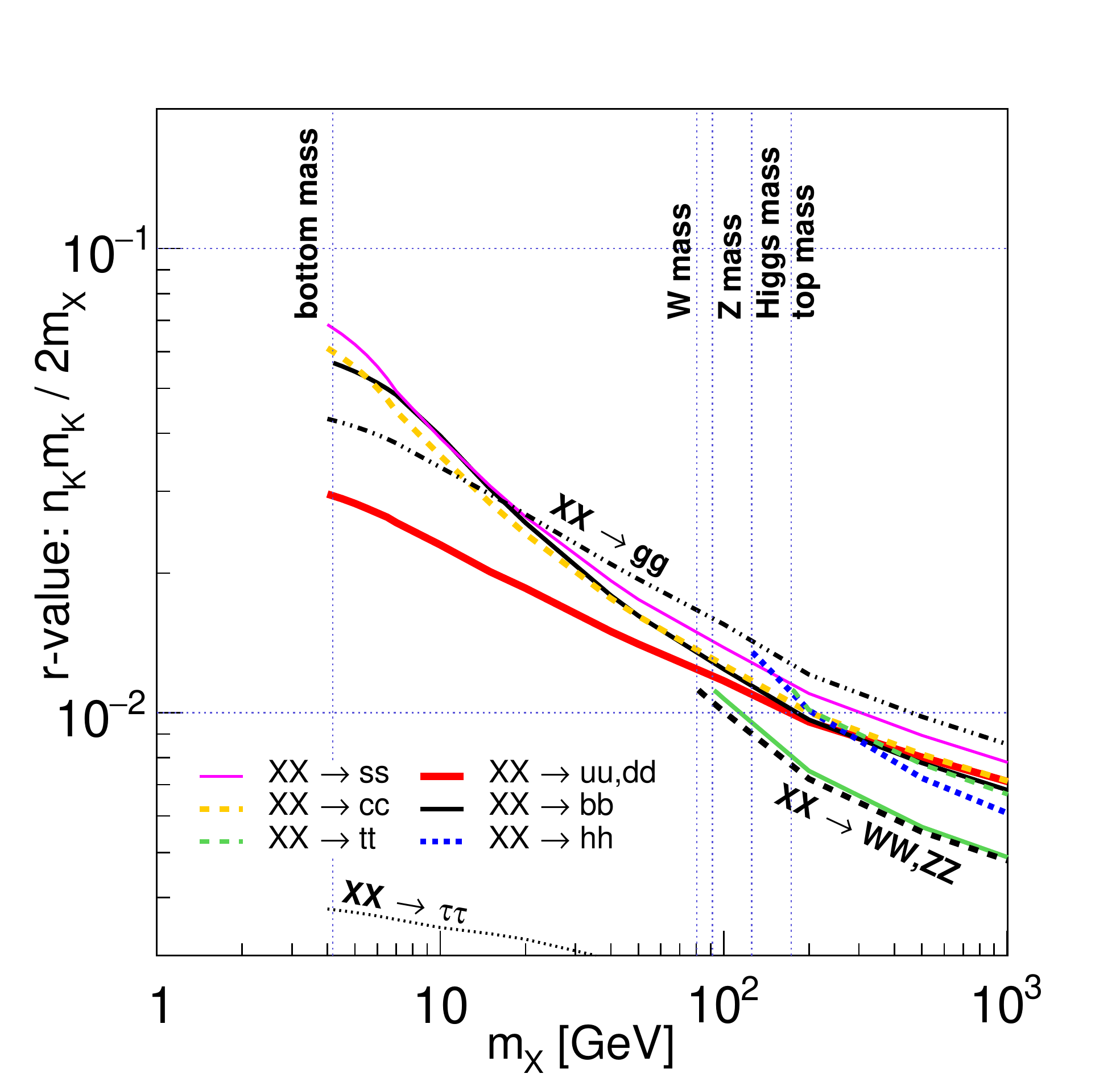}
  \caption{
  Fraction of the energy produced in dark-matter annihilation which is converted to stopped $\pi^+$ (left) and $K^+$ (right), $r_{\pi,K}$.
These fractions are then used to calculate the number of monoenergetic neutrinos produced in DM annihilation.  These fractions were calculated by simulating DM annihilation to hadrons in \textsc{Pythia}, then simulating the showering of the hadrons as they propagate through the solar medium using \textsc{GEANT}, as discussed in the text.}
  \label{fig:rvalues}
\end{figure}

The values of $r_{\pi}$ and $r_{K}$ as a function of dark matter mass are shown for the $\bar q q$ ($q=u,d,s,c,b, t$), $gg$, $\bar \tau \tau$, $WW$, $ZZ$, and $hh$ channels in Fig.~\ref{fig:rvalues}. Kaon decays at rest produce charged pions ($K^+ \rightarrow \pi^{+} \pi^{0},\pi^{+} \pi^{0} \pi^{0} ,  \pi^{+} \pi^{+} \pi^{-}$) with a branching fraction of about $28\%$; the stopped $\pi^+$ produced as a result of these decays are also included in $r_{\pi}$.
The $\bar uu$ or $\bar dd$ annihilation channels result in nearly the identical values for  $r_{\pi}$ and $r_{K}$ and are hence plotted as a single line.
For all final state channels except $\bar \tau \tau$, values of $r_{\pi}$ lie within roughly within a factor of 2 of each other for fixed $m_X$ (the
same is true for $r_{K}$).
This is because most pions and kaons are produced through hadronic cascade showers which have little dependence on the identity of the initial
particle which generated the shower.
Values of $r_{\pi ,K}$ are much lower for the $\bar \tau \tau$ channel, because a sizeable fraction of the initial energy
is always released as neutrinos.
It is interesting to note that $r_\pi$ is slightly greater for the $gg$ and $\bar s s$ channels than for the $\bar u u$ and $\bar d d$ channels.
The reason is because dark-matter annihilation to the $\bar u u$ and $\bar d d$ can initially produce a significant fraction of
$\pi^0$, which quickly decay to photons and produce no neutrinos.  Annihilation to the $\bar s s$ and $gg$ channels will tend to produce a slightly
smaller fraction of $\pi^0$ from hadronization.

The fraction of monoenergetic $\nu_\mu$ which oscillate to $\nu_e$ at 1~AU, including the effect of
oscillations in the Sun and vacuum, can be found in~\cite{Lehnert:2007fv}. We will use the values found
in~\cite{Lehnert:2007fv} assuming $\theta_{13} = 12^\circ$; this differs slightly from the current experimental
measurement, but the difference is not significant.  For a normal hierarchy, we have $F_{\nu_e}(E=30\mev) \sim 0.36$,
$F_{\nu_e}(E=236\mev) \sim 0.46$, while for an inverted hierarchy we have $F_{\nu_e}(E=30\mev) \sim 0.35$,
$F_{\nu_e}(E=236\mev) \sim 0.34$.

\subsection{Detector Effective Areas}

The relevant charged-current scattering processes at the detector are
$\nu_e + A \rightarrow e^- + A' +X$, where $A$ is a target nucleus, $A'$ is a final state nucleus, and $X$ refers to other
possible particles.  For LS, LArTPC and WC detectors, the relevant targets are $A={}^{12}\text{C}, {}^{40}\!\text{Ar}$ and ${}^{16}\text{O}$,
respectively.  In the case where $E_\nu = 236\mev$, these scattering cross sections can be evaluated using the \textsc{GENIE} software package~\cite{Andreopoulos:2009rq}.  For the 30 MeV neutrino arising from stopped $\pi^+$ decay, we will in this work only calculate sensitivities at LAr detectors; since 30 MeV is below the accepted range of validity of GENIE, we will instead use the scattering cross section against argon as calculated in~\cite{Kolbe:2003ys}.

The resulting cross sections are:
\bea
\sigma_{\nu_e + {}^{40}\!\text{Ar} \rightarrow e^- + A'+X } (30\mev) &\sim&  1.8 \times 10^{-40} \cm^2 ,
\nonumber\\
\sigma_{\nu_e + {}^{12}\text{C} \rightarrow e^- + A'+X } (236\mev) &\sim& 1.6 \times 10^{-38} \cm^2 ,
\nonumber\\
\sigma_{\nu_e + {}^{40}\!\text{Ar} \rightarrow e^- + A'+X } (236\mev) &\sim&  5.2 \times 10^{-38} \cm^2 ,
\nonumber\\
\sigma_{\nu_e + {}^{16}\text{O} \rightarrow e^- + A'+X } (236\mev) &\sim&  2.0 \times 10^{-38} \cm^2 .
\eea
For future liquid scintillation detectors where exposures may be much larger, one may need a careful calculation of the
cross section for charged-current scattering against carbon at 30 MeV (see, for example,~\cite{Auerbach:2001hz,SajjadAthar:2005ke}).

The effective area $A_{\text{eff}}$ of the detector can then be expressed as
\bea
A_{\text{eff}} &=& \sigma_{\nu_e -A} \times {M_{\text{target}} \over \kT} \times {(6.022 \times 10^{23}) \times 10^9 \over A} ,
\eea
where $A$ is the atomic mass of the target, $M_{\text{target}}$ is the fiducial mass of the target nucleus (in kT),
and $\sigma_{\nu_e -A}$ is the energy-dependent charged-current scattering cross section for an electron
neutrino and the target nucleus. To convert the detector fiducial volume to $M_{\text{target}}$, we apply factors of
$\sim 6/7$ for the carbon fraction in KamLAND and $\sim 8/9$ for the oxygen fraction of water.

\subsection{Backgrounds}

For LS and LArTPC detectors, the dominant background will arise from atmospheric electron neutrinos.  For the
energy range of interest, the angle-averaged atmospheric neutrino background can
be estimated~\cite{Battistoni:2005pd} as
\bea
{d^2 \Phi \over dE d\Omega} (E=30\mev) &\sim& 10 \m^{-2} \s^{-1} \sr^{-1}\mev^{-1} ,
\nonumber\\
{d^2 \Phi \over dE d\Omega} (E=236\mev) &\sim& 1 \m^{-2} \s^{-1} \sr^{-1}\mev^{-1} .
\eea
Atmospheric electron anti-neutrinos are also a background to a search for monoenergetic
neutrinos arising from stopped $\pi^+/K^+$ decay; conservatively, we assume that this will increase the background rate by
a factor of two.
Additionally, the background neutrino flux can vary from this estimate by a roughly a factor of two depending on the location of the detector~\cite{Battistoni:2005pd}.

At WC detectors the visible energy (energy of the final state electron) in the quasi-elastic scattering
process with oxygen ($\nu_e + {}^{16}\text{O} \rightarrow e^- + {}^{16}\text{F}$) is only about half of the initial neutrino
energy for the $30\mev$ line. At this energy range a series of other backgrounds besides
the atmospheric electron neutrino backgrounds needs to be considered.
Atmospheric $\bar \nu_\mu / \nu_\mu$ can, through a charged-current interaction in the
detector, produce a low-energy muon which then decays at rest yielding a Michel electron with energies below 52~MeV.
If the low-energy muon is below the Cherenkov threshold (sometimes called an \textit{invisible muon}), then it
would be difficult to distinguish this background from an electron-neutrino interaction.
Neutron tagging with gadolinium can reduce the invisible muon background with respect to electron \textit{anti}-neutrinos detected
via the inverse beta decay (IBD) reaction~\cite{Beacom:2003nk}, but is not applicable to the neutrino detection
channels used here. Other backgrounds arise from atmospheric neutrino neutral-current elastic events and
the production of charged pions in neutral-current reactions~\cite{Bays:2011si}. While the background estimation for
$30\mev$ line is complex, the target region is also subject to intense analysis
efforts for diffuse supernova neutrino searches~\cite{Beacom:2010kk}.
Far easier is the background estimation for the $236\mev$ line, which is dominated by the
atmospheric neutrino background~\cite{Takhistov:2014pfw}.  We will thus only calculate the sensitivities of WC detectors to the monoenergetic neutrino from kaon decay.

Prospects for line searches at both $30\mev$ and $236\mev$ are
good at LS or LArTPC detectors. Both LArTPC~\cite{Adams:2013qkq} and
LS~\cite{Learned:2009rv,Peltoniemi:2009xx} detectors are
expected to have very good track reconstruction and lepton discrimination capabilities. Since they do not rely on Cherenkov light as a detection mechanism,
invisible muons are not a concern. LS detectors can reconstruct the charged-lepton track
from the timing of when the photomultiplier
tubes (PMTs) are illuminated~\cite{Learned:2009rv,Peltoniemi:2009xx}.
A more complete analysis of background discrimination at LS and LArTPC detectors is beyond
the scope of this work.  We will assume, for simplicity, that the effects of the atmospheric $\nu_\mu$ background
are negligible at LS and LArTPC detectors.

Another potential source of background arises from cosmic rays which strike the Earth, Moon,
or Sun and produce $\pi^+$s and $K^+$s that are stopped and decay at rest. The number of stopped
mesons produced on Earth can be conservatively estimated by assuming that all cosmic ray pions that are produced
in the atmosphere and are boosted enough to not decay in flight reach the surface of the Earth
with no energy loss and convert 10\% of their energy into $\pi^+$s decaying at rest. The resulting
monoenergetic neutrino flux obtained with this estimate is about an order of magnitude smaller compared
to the atmospheric neutrino flux in the same energy range, assuming an energy resolution
$\epsilon \sim 5\%$.
For WC detectors, the larger value of $\epsilon \sim 15\%$ at $30\mev$ ($15\mev$ visible energy) would be
more appropriate~\cite{Fukuda:2002uc,Cravens:2008aa}.
This background is thus subleading, and may be
ignored.  The background arising from the generation of $K^+$ as pions stop in the Earth is similarly
negligible. On the Moon, due to the lack of an atmosphere, cosmic rays strike the lunar surface directly.
We assume 100\% of the energy of the cosmic rays is
converted to particles showers on the Moon and 10\% of this energy results in
$\pi^+$ and $K^+$ which decay at rest. We find that this background is also insignificant.
The background arising from cosmic rays striking the Sun is no greater than that arising from
cosmic rays striking the Moon, since they both have the same angular size when viewed
from the Earth, and can hence also be neglected. We can thus
ignore these backgrounds in the remainder of this analysis.

\subsection{Energy Resolution}

For a 30 MeV neutrino, the dominant charged-current processes are $\nu_e + A \rightarrow e^- + A'$. The electron is thus
monoenergetic in center-of-mass frame and its energy, $E_e^{\text{cm}}$, is entirely determined by the neutrino energy:
\bea
E_e^{\text{cm}} &=& {E_\nu - (m_{A'}^2 - m_A^2 -m_e^2)/(2m_A) \over  \sqrt{1+ 2(E_\nu / m_A) } } \sim E_\nu - \Delta m ,
\eea
where $\Delta m = m_{A'} - m_A$.
Thus, a measurement of $E_e^{\text{cm}}$ is equivalent to a measurement of the neutrino energy.
For $A={}^{40}\!\text{Ar}$, $\Delta m \sim 1.5\mev$, and $E_e^{\text{cm}} \sim E_\nu$.  But for
$A={}^{12}\text{C}, {}^{16}\text{O}$, we find $\Delta m \sim 17.5\mev, 15.5\mev$, respectively.  For these targets, if $E_\nu = 30\mev$,
then the electron energy is significantly different from the neutrino energy.  The relevant backgrounds
are then those which would produce an electron of the appropriate energy.  On the other hand, one might also be
able to measure the energy released when the excited final state nucleus $A'$ decays.  For example, ${}^{12}\!\text{N}$ will
$\beta$-decay on a time-scale of approximately $16~{\rm ms}$~\cite{Auerbach:2001hz};
this process may allow one to better reconstruct the neutrino
energy and reject backgrounds at an LS detector.
For these cases, the background
analysis is more difficult; we will therefore not present sensitivities for LS or WC detectors for the $\pi^+$ channel.

The non-relativistic boost from center-of-mass frame to the
detector frame will smear out this monoenergetic signal, resulting in a full-width $\Delta E$ given by
\bea
\Delta E &\sim& \left({2 E_\nu \over m_A} \right) E_e^{\text{cm}}  .
\eea

One can estimate the energy
resolution for electron reconstruction for
LS~\cite{Peltoniemi:2009xx}, LArTPC~\cite{Adams:2013qkq} and WC~\cite{Ashie:2005ik,Evslin:2015pya} detectors in this energy range.
For LS and LArTPC detectors, a reasonable estimate would be $5\%$.  For WC detectors, one could estimate the energy resolution
for electron reconstruction as $60\% (E_e /\mev)^{-1/2}$; this gives $\sim 15\%$ for a $\sim 15\mev$ electron, and $\sim 4\%$ for a $\sim 215\mev$ electron.

For a 236 MeV neutrino, a quasielastic scatter can also liberate a proton from the target nucleus.  For an LS or LArTPC
detector, the neutrino energy resolution would not suffer dramatically since the energy of the released proton can be
well measured.  For WC detectors, however, it would be more difficult to reconstruct the neutrino energy for such events.

Accounting for these contributions, we will make a simple benchmark estimate for the neutrino energy resolution of $\epsilon =10\%$
for $236\mev$ neutrinos, as well as for $30\mev$ neutrinos at LArTPC detectors.  Although this is a reasonable estimate,
a more accurate assessment of the energy resolution would require a more detailed analysis which is beyond the scope of this
work.  However, the sensitivities found here can be rescaled for any choice of the energy resolution.  In the limit of large
background, the sensitivity scales inversely with the square root of the energy resolution ($\sigmaSD^p \propto \epsilon^{-1/2}$),
while it is independent of the energy resolution in limit of small background.

\section{Results}

The number of signal ($N_S$) and background ($N_B$) events which can be observed at the detector, within an
energy window $(1 \pm \epsilon/2)E_0$, can be expressed as
\bea
N_{S,B} &=& \left[f_{S,B} \int_{(1 - \epsilon/2)E_0}^{(1 + \epsilon/2)E_0} dE \int d\Omega
{d^2 \Phi_{S,B} \over dE d\Omega} \right] \times A_{\text{eff}} \times T ,
\eea
where $T$ is the time exposure, $\Phi_{S,B}$ is either the signal or background flux,
and $f_{S,B}$ is the fraction of events that fall within the neutrino
energy bin centered at 29.8~MeV or 235.6~MeV (depending on whether we are finding the number of events from the pion or kaon monoenergetic neutrino, respectively). If $\Phi_B$ is a relatively smooth background flux, as is
the case here, then $f_B=1$.
If $N_B \ll 1$, then one may choose $\epsilon$ to be larger than the energy resolution without greatly increasing the
number of background events; most of the monoenergetic neutrino signal events will thus fall
within the energy bin, implying $f_S \sim 1$.
But if the number of background events is not small and one chooses $\epsilon$ to be given
by the fractional energy resolution of the detector, then $f_S \sim 0.68$, as only
$\sim 68\%$ of the monoenergetic events will be reconstructed within the chosen energy bin.
Note that the angular integrals cover all directions, because neutrinos of such low energies
will produce a largely isotropic distribution of charged leptons.

We will derive the 90\% CL sensitivity of benchmark LS (KamLAND), LArTPC (DUNE) and WC (Super-Kamiokande, Hyper-Kamiokande)
detectors in the $(m_X, \sigmaSD^p)$-plane, in
the case where dark matter annihilates exclusively to first generation quarks,\footnote{As
$N_S \propto \sigmaSD^p \times r_{\pi,K} \times F_{\nu_e}$, the sensitivity in any other channel or with any choice of
hierarchy is determined by rescaling the given sensitivity by
the ratio of $r_{\pi,K}$ values given in Fig.~\ref{fig:rvalues}, or by the appropriate ratio of $F_{\nu_e}$ values.}
assuming a search for monoenergetic neutrinos at either $30\mev$ (stopped $\pi^+$ decay) or
$236\mev$ (stopped $K^+$ decay).
For Super-Kamiokande, we assume a fiducial volume of 22.5 kT with a runtime of 3903 days~\cite{Choi:2015ara}.
For a search for fully-contained $\nu_e$ events at KamLAND, we assume a fiducial volume of 0.4 kT~\cite{Kumar:2009ws,Kumar:2011hi},
with a runtime of $\sim 3600$ days~\cite{KamLAND}.
We will assume a normal hierarchy.
To derive a sensitivity, we will assume that the number of observed events is
set by the expected number of events due to background.  In Table~\ref{Tab:Experiments} we list, for each experiment and channel,
the exposure,
expected number of background events, assumed number of observed events, fraction $f_S$ of signal events falling within an energy bin, and the minimum number of expected signal events such that (given the number of assumed observed events) a model would be excluded at 90\% CL.

\begin{center}
\begin{table}[h]
\begin{tabular}{|c|c|c||c|c|c|c||c|c|c|c|}
  \hline
  experiment  & status & exposure & $N_B^{\pi}$ & $N_{\text{obs}}^{\pi}$ & $f_S^{\pi}$ & $N_S^{\pi}$ & $N_B^{K}$ & $N_{\text{obs}}^{K}$ & $f_S^{K}$ & $N_S^{K}$   \\
\hline
  KamLAND & current& 4~kT~yr & --- & --- & --- & --- & 5.1 & 6 & 0.68 & 5.5 \\
  DUNE & future & 34~kT~yr & 0.2 &  0 & 1 & 2.3 & 50 & 50 & 0.68 & 10.3 \\
  Super-K & current & 240~kT~yr & --- & --- & --- & --- & 305 & 305 & 0.68 & 28.7 \\
  Hyper-K & future & 600~kT~yr & --- & --- & --- & --- & 762.5 & 763 & 0.68 & 45.4 \\
  \hline
\end{tabular}
\caption{For each experiment we list the assumed exposure in kT yr.  We also list (for $\pi^+$ and $K^+$ channels), the expected
number of background events ($N_B$), the assumed number of observed events ($N_{\text{obs}}$), the fraction of signal events which fall within
the energy bin ($f_S$), and the minimum number of expected signal events ($N_S$) such that a model would be excluded at 90\% CL, given the assumed number of observed events.}
\label{Tab:Experiments}
\end{table}
\end{center}

\begin{figure}[t]
  \includegraphics[scale=0.7]{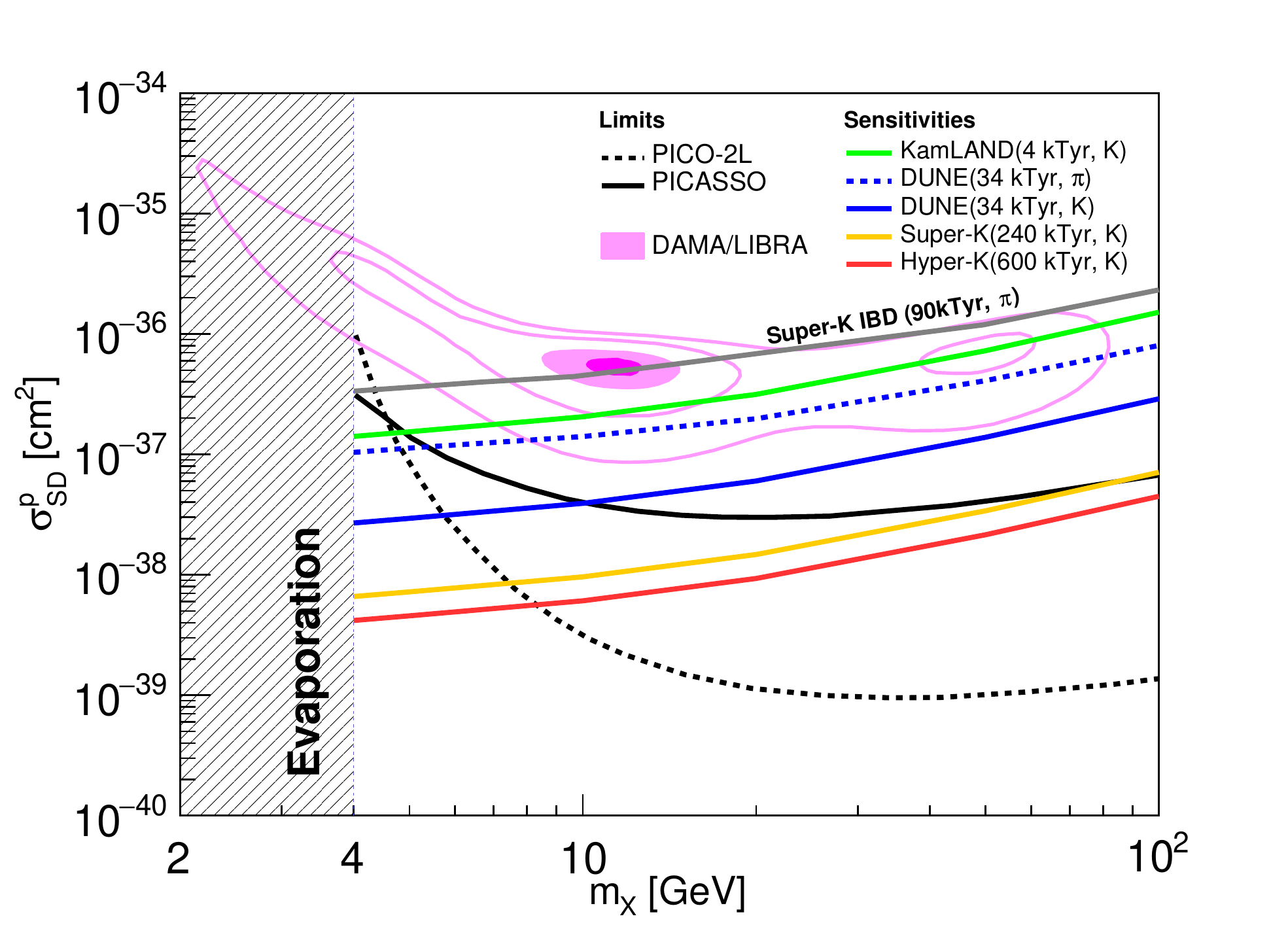}
  \caption{The 90\% CL sensitivity curves for (in order of increasing sensitivity) KamLAND (4~kT~yr), DUNE (34~kT~yr), Super-Kamiokande (240~kT~yr)
  and Hyper-Kamiokande (600~kt~yr) in the stopped $K^+$ (solid) channel.  For DUNE also the stopped $\pi^+$ (dotted) channel is shown.
  We assume $\epsilon = 10\%$, a normal hierarchy, and that dark matter in the
  Sun annihilates only to first generation quarks.  Also plotted are the 90\% CL bounds from PICASSO~\cite{Archambault:2012pm}
  and PICO-2L~\cite{Amole:2015lsj} (dashed grey and black, respectively).  The shaded area represents the region where interactions with the thermal bath of nucleons in the Sun cause dark matter particles to easily be ejected (or evaporate) from the Sun~\cite{WIMPevaporation}. Scenarios motivated by the DAMA/LIBRA annual modulation signal~\cite{Savage:2008er} are also shown.  The dark (light) solid regions correspond to
  90\% ($3\sigma$) CL, and the two outer contours are $5\sigma$ and $7\sigma$ CL.  Results are compared to sensitivities derived previously~\cite{Rott:2012qb,Bernal:2012qh}  with 4~years of Super-Kamiokande data (or 90~kT~yr) in the inverse beta decay (IBD) channel.}
  \label{fig:sensitivity}
\end{figure}

In Fig.~\ref{fig:sensitivity} we plot the 90\% CL sensitivities for dark matter annihilation into light quarks in the
stopped $\pi^+/K^+$ decay channels of KamLAND,
DUNE, Super-Kamiokande and Hyper-Kamiokande, as well as the 90\% CL limits of PICASSO~\cite{Archambault:2012pm} and PICO-2L~\cite{Amole:2015lsj}.
Note, we do not plot the sensitivity of KamLAND, Super-Kamiokande or Hyper-Kamiokande to monoenergetic $30\mev$ neutrinos, since for
these targets the energy of the produced electron is relatively small compared to that of the incoming neutrino, making
the backgrounds harder to estimate.  In
any case, the sensitivity of KamLAND would, at best, be exceeded by current bounds
from PICASSO and PICO-2L.
At low mass ($m_X \sim 5\gev$),
current data from KamLAND (in the $K^+$ channel) can provide limits which are competitive with the best direct-detection
limits from PICO-2L.  DUNE's sensitivity in the $K^+$ channel will be even better with
$34 \kT \yr$ exposure.  However, the sensitivity of Super-Kamiokande, available with current data,
will exceed that of DUNE and KamLAND as a result of Super-Kamiokande's large exposure.
The proposed searches can also test dark matter scenarios motivated by the DAMA/LIBRA annual modulation signal~\cite{Savage:2008er};
the DAMA/LIBRA signal region is also shown in Fig.~\ref{fig:sensitivity}.
Our sensitivities shown are for dark matter annihilations into light quarks, which would not yield any significant number high energy
neutrinos and hence not be detectable via traditional searches at neutrino telescopes. Our sensitivities can easily be
rescaled to any other annihilation channel by taking the ratio of the $r$-values (shown in Fig.~\ref{fig:rvalues}) between the $uu,dd$-channel with
the channel of interest.

We compare our results to earlier results derived with 4 years of Super-Kamiokande data (or 90~kT~yr) in the
inverse beta decay channel~\cite{Rott:2012qb,Bernal:2012qh}, which are also plotted in Fig.~\ref{fig:sensitivity}.
Compared with our estimate of the Super-Kamiokande kaon signal, an improvement
of roughly 2~orders of magnitude can be seen. The origin of the improved sensitivity for Super-Kamiokande
with our proposed monoenergetic neutrino channel from kaon decay at rest over the
continuum anti-neutrino channel arising from pion decay at rest can be understood as follows: foremost, neutrino cross sections
in the MeV range rise quadratically with energy, which makes the kaon line more detectable even
though the kaon yield from the hadronic showers in the Sun is about an order of magnitude smaller than the pion yield.
Roughly another factor of two is gained by more favourable neutrino oscillation effects; while the inverse beta decay channels
utilize the 1/6 of muon anti-neutrinos that have oscillated to electron anti-neutrinos at the detector, the kaon line is detected
via the 1/3 of muon neutrinos that have oscillated to electron neutrinos.
The kaon channel further benefits from lower atmospheric and detector backgrounds; further the
spectral feature of the line signature allows for a narrow energy window.
All these features make the kaon line neutrino channel the most promising detection channel
for signals of dark matter annihilation in the Sun into light quarks. Only in a nearly background
free environment of a gadolinium doped WC detector would the
sensitivity of the electron anti-neutrino IBD channel from pion decay
at rest become comparable~\cite{Rott:2012qb,Bernal:2012qh}.

One should note that DUNE is largely signal-limited in the $\pi^+$ channel,
\textit{i.e.,} at the limit of their sensitivity, $S/B \gtrsim 1$.  The reason is that the neutrino--nucleus
charged-current scattering cross section drops drastically with decreasing neutrino energy.
As a result, the sensitivity is limited only by the exposure needed to obtain a few
signal events, and is largely independent of the energy resolution.

The $K^+$ channel is much less signal-limited; because the neutrino nucleus scattering cross section is more than two orders of
magnitude larger at $E= 236\mev$ than at $E=30\mev$, much less exposure is needed for a given dark matter model
in order to obtain a few signal events in the $K^+$ channel, as compared to the $\pi^+$ channel.
But on the other hand, the monoenergetic $\pi^+$ channel has a much larger signal-to-background ratio than the
$K^+$ channel.  Although the atmospheric $\bar \nu_e/\nu_e$ background flux is an order of magnitude smaller at
$E=236\mev$ than at $30\mev$, this is compensated by the larger energy bin size at $E=236\mev$.  Moreover, the
number of $30\mev$ neutrinos produced per annihilation is typically more than
an order of magnitude larger than the number of $236\mev$ neutrinos.
While the $K^+$ channel will provide better sensitivity
with a fixed exposure, the monoenergetic $\pi^+$ channel provides the better signal-to-background ratio.

We can compare the sensitivities from LS and LArTPC detectors with those expected from water Cherenkov detectors.
While LS and LArTPC detectors may outperform these WC detectors in background rejection, water Cherenkov
detectors are easier to realize with a large detector volume.  For 236 MeV neutrinos, the sensitivity of DUNE,
Super-Kamiokande and Hyper-Kamiokande would scale as $\sigmaSD^p \propto \epsilon^{-1/2}$, while for KamLAND the
scaling of sensitivity with energy resolution is more complicated because the number of background events is neither
large nor negligible.
The different energy ranges for neutrinos arising from pion decay and kaon decay at rest have been surveyed before in context
of the search for diffuse supernova neutrino background~\cite{Bays:2011si} and in the context of atmospheric neutrino oscillation
measurements~\cite{Ashie:2005ik}, respectively.
Competitive searches could also be carried out by JUNO~\cite{An:2015jdp}, RENO50~\cite{Kim:2015dag}, or at
large underground detectors in the JinPing laboratory.
We note that the kaon line might in the future also be accessible at large volume neutrino telescopes.
Augmentations to PINGU~\cite{Aartsen:2014oha} or ORCA~\cite{Katz:2014tta}
could give access to neutrinos at 236~MeV.

\section{Conclusion}

We have considered the sensitivity of neutrino detectors to dark matter which
annihilates in the Sun to light quarks.  The result of dark-matter annihilation
in these models is a shower of light hadrons, producing an abundance of stopped
$\pi^+$ and $K^+$ whose decays yield monoenergetic neutrinos.  The energy of
monoenergetic electron neutrinos can be well measured at LS and LArTPC detectors,
allowing one to search for a line signal with a dramatically reduced background.
An analysis of current KamLAND data already provides limits on this class of models
which are competitive with current limits from PICO-2L and PICASSO.
Sensitivity can be even better with 34~kT~yr of data from DUNE, or with
an analysis of current data from Super-Kamiokande.

The most promising models, in terms of sensitivity, are those with low dark matter
mass ($\lesssim 10\gev$).  This region of parameter space is difficult to probe with
direct-detection experiments, which tend to lose sensitivity rapidly at low mass.
 If dark matter annihilates primarily to light quarks, then more traditional searches for
hard neutrinos emanating from the Sun will also be ineffective.  As a result, searches
for the neutrinos arising from stopped meson decay are a very effective and complementary search
strategy.

This strategy is characterized by a very large signal-to-background ratio.  The downside,
however, is that a very large exposure is required to fully exploit this strategy, due to
the small neutrino--nucleus scattering cross section at these energies.  Future large exposure
detectors will be particularly well-suited for this detection strategy.  It is interesting to
compare the sensitivities of different detector types (LS, LArTPC, WC) in the $\pi^+$ and $K^+$
channel.  The $\pi^+$ channel produces a better signal-to-background ratio than the
$K^+$ channel for a given model, but this channel is signal-limited unless the detector has a
very large exposure; for example, DUNE (34 kT) would require about 5 years of running to fully exploit
this channel.  WC detectors can more easily be produced with very large exposures, but suffer from
much poorer background rejection in this channel.  WC detectors with a much larger exposure may be easier to build, but sensitivity will only
increase as the square root of the increase in exposure.  The sensitivity of LS and LArTPC detectors will continue
to grow linearly with exposure, provided such large exposures can be
realized.   In the $K^+$ channel, however, DUNE can approach sensitivity in the $S/B \sim 1$ regime with only
months of runtime. A positive detection would benefit from the possibility of independently verifying
a signal in different detectors and by different channels, ranging from the neutrino line channels
at $29.8\mev$ and $236\mev$, to the anti-neutrino continuum via inverse beta decay.

The largest source of uncertainty in the reported sensitivities originates from the neutrino
cross sections in the MeV range. In the future we can expect that beam dump
experiments as a source of stopped pions will result in
reliable cross section measurements and hence can be used to obtain
more robust sensitivity estimates.  An example is the
${\rm DAE}\delta{\rm ALUS}$ proposal~\cite{Conrad:2010eu}, for which stopped
pions would be produced by a proton
beam impinging on a fixed target in the vicinity of DUNE.  
Stopped kaons could also be produced by such experiments~\cite{Spitz:2012gp}.
The neutrinos from such an experiment would provide an excellent calibration source for the
proposed search for dark matter utilizing neutrinos from the Sun. 
Due to the time structure of
the proposed source, with 60~Hz prompt muon neutrinos and a
delayed signal with time scale of $2.2~\mu{\rm s}$ of anti-muon and electron neutrinos, it would
not contribute in any significant way to the background of the monoenergetic neutrino search.

Although we have considered the case where monoenergetic $\nu_\mu$s produced at the core of the Sun oscillate
to $\nu_e$s by the time they reach the detector, one could also consider the case where the neutrino reaching
the detector is still a $\nu_\mu$.  For a $30\mev$ neutrino, this channel would be ineffective since the neutrino
does not have enough energy for a charged-current interaction.
A 236 MeV $\nu_\mu$, however, could produce a low-energy $\mu^-$ through a charged-current interaction at the detector 
(this process was considered in~\cite{Spitz:2014hwa}).
This muon would be invisible at a WC detector, since it quickly falls below its Cherenkov threshold.  But it could be
observed at an LS or LArTPC detector, potentially providing another channel by which these detectors could
probe models in which dark matter annihilates to light quarks in the core of the Sun.

\vskip .2in
\acknowledgments

We are grateful to Prateek Agrawal, Erin Edkins, Yu Gao, Jennifer Gaskins, Bira van Kolck, John G.~Learned,
Danny Marfatia, Sandip Pakvasa, Michinari Sakai, and especially John Beacom and Shao-Feng Ge for useful discussions.  The work of J.~Kumar is
supported in part by NSF CAREER Grant No.~PHY-1250573.  C.~Rott is supported by the Basic Science Research Program
through the National Research Foundation of Korea funded by the Ministry of Science, ICT $\&$ Future planning NRF-2013R1A1A1007068.
S.~In is supported by Global PH.D Fellowship Program through the National Research Foundation of Korea (NRF) funded by the Ministry of Education (NRF-2015H1A2A1032363). D.~Yaylali is supported in part by DOE grant DE-FG02-13ER-41976.
J.~Kumar and C.~Rott would like to thank the Aspen Center for Physics and CETUP* (Center for Theoretical
Underground Physics and Related Areas), for hospitality and partial support.





\end{document}